\documentclass[sigconf]{acmart}
\acmConference[CAIN 2024]{3rd International Conference on AI Engineering — Software Engineering for AI}{April 2024}{Lisbon, Portugal}

\usepackage{listings}
\usepackage{xcolor}

\usepackage{algorithmic}
\usepackage{framed}
\usepackage{mdframed}
\usepackage{textcomp}
\usepackage{rotating}
\usepackage{caption}
\usepackage{subcaption}
\usepackage{hyperref}
\usepackage{textcomp}
\usepackage{tikz}
\usepackage{multirow}
\newcommand{\numberedcircle}[1]{%
  \begin{tikzpicture}[baseline=(char.base)]
    \node[shape=circle, draw, inner sep=1pt] (char) {\small #1};
  \end{tikzpicture}}
\def\BibTeX{{\rm B\kern-.05em{\sc i\kern-.025em b}\kern-.08em
    T\kern-.1667em\lower.7ex\hbox{E}\kern-.125emX}}

\AtBeginDocument{%
  \providecommand\BibTeX{{%
    \normalfont B\kern-0.5em{\scshape i\kern-0.25em b}\kern-0.8em\TeX}}}

\newcommand{\toolname}{GreenRunner}

\begin{document}

\title{Green Runner: A tool for efficient deep learning component selection}


\author{Jai Kannan, Scott Barnett 
}

\email{{jai.kannan, scott.barnett}@deakin.edu.au}
\affiliation{%
  \institution{Applied Artificial Intelligence Institute}
  \streetaddress{Deakin University}
  \city{Geelong}
  \country{Australia}
}

\author{Anj Simmons, Taylan Selvi 
}

\email{{a.simmons, taylan.selvi}@deakin.edu.au}
\affiliation{%
  \institution{Applied Artificial Intelligence Institute}
  \streetaddress{Deakin University}
  \city{Geelong}
  \country{Australia}
}

\author{Luís Cruz 
}

\email{l.cruz@tudelft.nl}
\affiliation{%
  \institution{Delft University of Technology}
  \streetaddress{Delft}
  \city{Delft}
  \country{Netherlands}
}


\begin{abstract}

For software that relies on machine-learned functionality, model selection is key to finding the right model for the task with desired performance characteristics. Evaluating a model requires developers to i) select from many models (e.g. the Hugging face model repository), ii) select evaluation metrics and training strategy, and iii) tailor trade-offs based on the problem domain. However, current evaluation approaches are either ad-hoc resulting in sub-optimal model selection or brute force leading to wasted compute. In this work, we present \toolname, a novel tool to automatically select and evaluate models based on the application scenario provided in natural language. We leverage the reasoning capabilities of large language models to propose a training strategy and extract desired trade-offs from a problem description. \toolname~features a resource-efficient experimentation engine that integrates constraints and trade-offs based on the problem into the model selection process. Our preliminary evaluation demonstrates that \toolname{}  is both efficient and accurate compared to ad-hoc evaluations and brute force. This work presents an important step toward energy-efficient tools to help reduce the environmental impact caused by the growing demand for software with machine-learned functionality. Our tool is available at Figshare \href{https://figshare.com/s/248381647619ba223334}{GreenRunner}. 
\end{abstract}



\maketitle

\begin{figure*}[tbp]
    \centering
    \includegraphics[width= \linewidth]{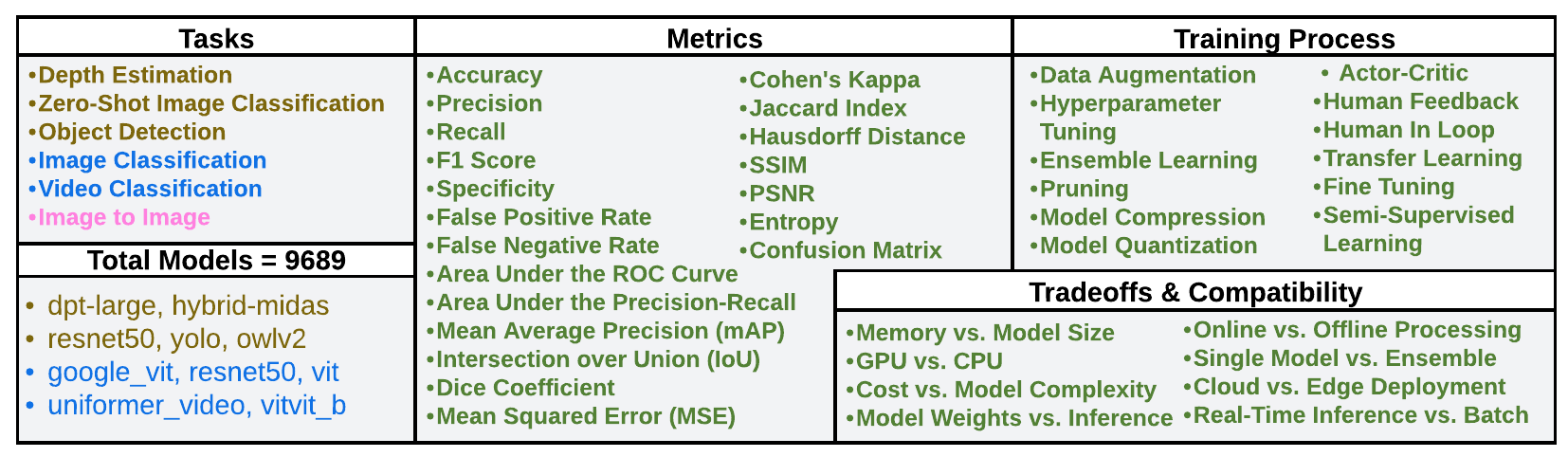}
    \caption{The problem space where developers have to pick $X$ tasks containing $N$ models, evaluate them with $E$ metrics and $T$ tradeoffs which needs to be evaluated iteratively, resulting in wasted resources and increased costs. }
    \label{fig:problemspace}
\end{figure*}
\section{Introduction}

While deep learning (DL) significantly enhances software, it also brings various challenges that need careful attention. A central concern is that of the environmental impact of deep learning models \cite{strubell2019energy, dodge2022measuring,zhou2023opportunities, luo2023achieving}. The environmental impact of training DL models has primarly been the focus \cite{ligozat2021unraveling, wu2022sustainable}. However, environmental impact that occurs when reusing deep learning models through additional training and evaluation of models is less studied. Selecting an appropriate deep learning component (a pretrained model or an API web service) requires a) evaluating against a large sample dataset, b) comparing multiple alternatives, and c) discovering an optimal reuse strategy (i.e. fine tuning, transfer learning, ensembling etc.) All this evaluation wastes compute resources. Finding energy efficient strategies to compare and select DL components is important to mitigate the environmental impact of deep learning.

Efficient approaches to select a DL component include 1) relying on recorded metrics on a benchmark dataset~\cite{kornblith2019better, barbu2019objectnet}, 2) using proxy metrics that approximate the actual performance~\cite{dwivedi2019representation,zhang2023model}, or 3) transferability estimation~\cite{bao2019information, tran2019transferability, bolya2021scalable, huang2022frustratingly}. These approaches aim to approximate the actual performance of a DL component to streamline evaluations. However, all approaches rely on approximating model performance neglecting crucial aspects like memory requirements, computational demands, and hardware compatibility. This narrow focus leads to suboptimal selection that does not align to a specific application context. The gap addressed in this paper is that existing DL component selection approaches do not consider the tradeoffs faced by software engineers when considering an application specific context. 

To illustrate tradeoffs that arise when selecting DL components consider the following examples. An agricultural drone to detect weeds requires a balance between prediction accuracy and energy efficiency to extended flight times. In contrast, an autonomous vehicle has less concerns around energy efficiency (i.e. they have larger batteries) but has greater demand for low-latency detection to ensure occupant safety~\cite{inbook}. The same DL task, detecting objects in images, produces different tradeoffs depending on the a) deployment scenario, b) safety criticality of the system, and c) hardware requirements. As a consequence, the application context where the DL component is going to be used influences the criteria used to make a selection.


Software engineers have to select more than a DL component (i.e. a model on Huggingface) but also the configurations for evaluation. A selection of tasks, models, metrics, reuse process and tradeoffs to consider are depicted in \autoref{fig:problemspace}, which presents only the computer-vision models on Huggingface\footnote{\url{https://huggingface.co/}}.  For image object detection, a developer must select from 1245 models\footnote{\url{https://huggingface.co/models?pipeline_tag=object-detection&sort=trending}}, select the right combination of metrics and training process and iteratively evaluate the models and then compute the tradeoffs. This highlights the need for a resource-efficient and cost-effective DL component selection strategy that considers 1)  operational criteria, 2) environmental impact, and 3) comparisons between competing components.

To assist software engineers in selecting deep learning (DL) components, we introduce \toolname. This tool streamlines the evaluation process by incorporating an energy-efficient experimentation engine based on a multi-armed bandit framework and utilizing the reasoning capabilities of a large language model (LLM). The LLM suggests application-specific configurations, including the model, metrics, reuse strategies, and trade-offs. Engineers input a description of their application and its operational context, and \toolname{} generates the necessary evaluation configuration.

Our approach is grounded in the hypothesis that an LLM, trained on scientific publications, blog articles, and open-source machine learning repositories \cite{chen2022towards, yang2023large, cummins2023large}, can produce near-optimal configurations. We posit that the computational cost of training and operating a single LLM is offset by the reduced compute requirements for future applications developed using DL components.

We demonstrate the effectiveness of our approach with a preliminary evaluation using the ObjectNet dataset and 71 object detection models, showcasing \toolname's potential in optimizing DL component selection.

\textbf{Key Contributions arising from this work:}
\begin{itemize}
    \item A preliminary empirical evaluation of an LLM's world model for informing component selection (focusing on DL component selection).
    \item An approach for optimising the DL component selection process. Our approach uses a 1) dynamically created multi-objective function derived from a natural language description of the problem, and 2) an efficient evaluation process based on multi-armed bandits. 
    \item A tool implementation of the approach, \toolname{}, available for download and \footnote{\url{https://figshare.com/s/248381647619ba223334} (anonymised for blind review)}is hosted at:https://green-runner-web-eehhqvzsqa-km.a.run.app/
\end{itemize}

\section{Motivating Example}\label{sec:motivation}

To motivate the need for \toolname, consider Tom, a software engineer at an agricultural drone manufacturing company. The WeedWhack drones are to detect and then spray weeds without affecting the main crop. Tom is tasked with implementing the weed detection component. Tom has been loosely following the advances in object detection but is primarily a software engineer, not a machine learning expert. To complete the task Tom needs to select an appropriate DL component by 1) finding a set of candidate components (i.e. online models or web services), 2) select evaluation metrics, reuse strategy, and hyperparameters, and 3) evaluate and compare the results. These tasks have to be completed while considering the constraints of the problem domain namely a) resource efficient models (cover the whole field without recharging), b) accurate detection (only target weeds not crops), and c) adapt to a multitude of weed types. 

Tom starts identifying DL components for WeedWhack by selecting object detection models from Huggingface and online services (e.g. AWS Rekognition API\footnote{\url{https://docs.aws.amazon.com/rekognition/latest/dg/API\_Reference.html}}). Tom selects a handful of components to evaluate against the dataset of weeds and starts evaluating them all against the benchmark dataset. Each component needs to be adapted to the weed dataset to find the ideal model which wastes time, increases compute costs, and is energy inefficient. Tom is also unaware of the range of strategies suitable for the problem domain (i.e. pruning, quantization, fine-tuning, and/or transfer learning) and ends up making a sub-optimal choice. Once the drone has been deployed Tom finds the model he selected sprays too much of the crop and needs to evaluate a new set of models. This wastes development time and increases the cost of implementation. 

What Tom needs is an approach that recommends the right DL component and strategy that considers the requirements of WeedWhack. Once recommendations have been made Tom needs an efficient process to compare competing strategies to make a timely yet adequate selection of a DL component.

\section{\toolname} 
\label{sec:solution-direction}

\begin{figure*}[h]
    \centering
    \includegraphics[width=\linewidth]{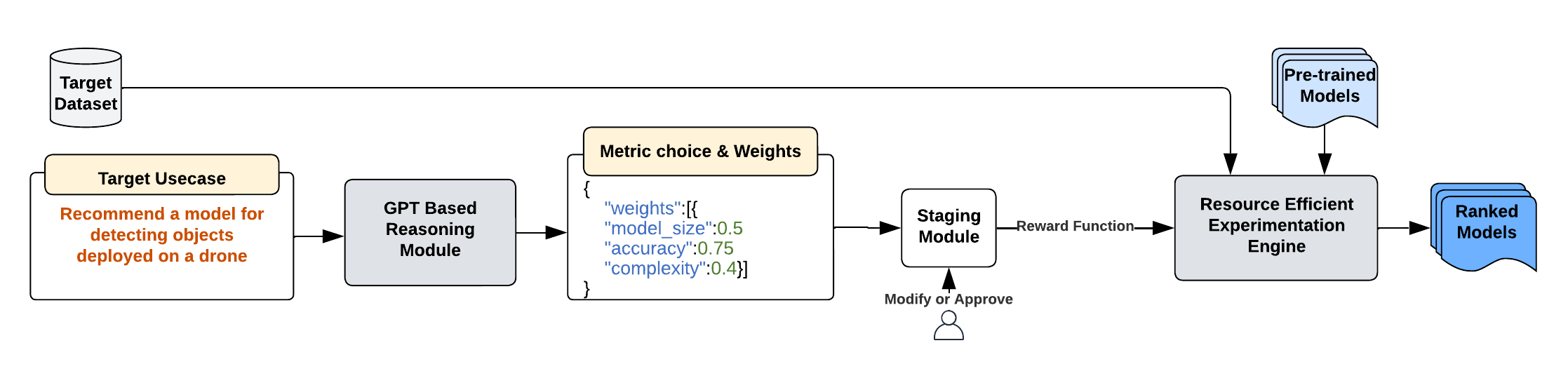}
    \caption{Overview of \toolname{} describing the internal processes and outputs from each process.}
    \label{fig:overview}
\end{figure*}

To design \toolname{} we took inspiration from the tool proposed by \citet{cummaudo2020threshy} , we describe the core components of \toolname, shown in figure \autoref{fig:overview}. The core components are 1) a GPT Based Reasoning Module to suggest metric choice \& weights for a target use case, which forms a reward function, and 2) a resource-efficient experimentation Engine that evaluates pre-trained models on a target dataset against using a reward function to produce a set of top-ranked components for the target use case.

\subsection{GPT Based Reasoning Module}
\label{sec:reasonmodule}

The integration of \toolname{} with an LLM enables users to provide a concise plain text prompt that describes the specific use case for the application which is the first step in identifying an optimal DL component for a particular use case. For this paper, the LLM used by the reasoning module was GPT-4. Analysing the description, the LLM generates a set of metrics and weights, which are used as part of a reward function for performing the selection process. These weights are distributed across the metrics suggested by the LLM such as: i) Model accuracy, ii) Model size, and iii) Model complexity.

Additionally, the LLM offers comprehensive justifications for the selected weights, optimizing them in accordance with the use case at hand. These metric weights serve as configuration parameters for resource-efficient evaluation algorithms, which effectively navigate the extensive model repository and select the most appropriate component. Prior to conducting the experiment, the staging module allows users to thoroughly examine and refine the metric weights, ensuring alignment with the intended use case.

\subsection{Resource Efficient Experimentation Engine}



\toolname{} utilizes multi-armed bandit (MAB) algorithms \cite{slivkins2019introduction} to streamline the selection of deep learning components for specific applications, reducing the need for extensive evaluations. These algorithms adeptly balance the exploration of diverse options with the exploitation of known advantageous actions.

In \toolname, deep learning models from a repository function as "arms". The system is designed to identify top-performing models with a limited number of evaluations. It employs a custom reward function that considers metrics such as accuracy, size, complexity generated by the reasoning module \autoref{sec:reasonmodule}, thereby adapting the selection to the unique requirements of each use case.

\toolname{} offers three MAB selection strategies: Epsilon Greedy, Upper Confidence Bound, and Thompson Sampling. The evaluation of these models is based on a user-defined budget,and strategy where a larger budget enhances the likelihood of finding the optimal model, whereas a smaller budget limits evaluations, potentially leading to the selection of less-than-ideal models. As more data is processed, the MAB algorithm continuously refines its understanding of each model's performance, effectively pinpointing the most suitable models for the specific dataset.




\begin{figure*}[h]
    \centering
    \includegraphics[width=\linewidth]{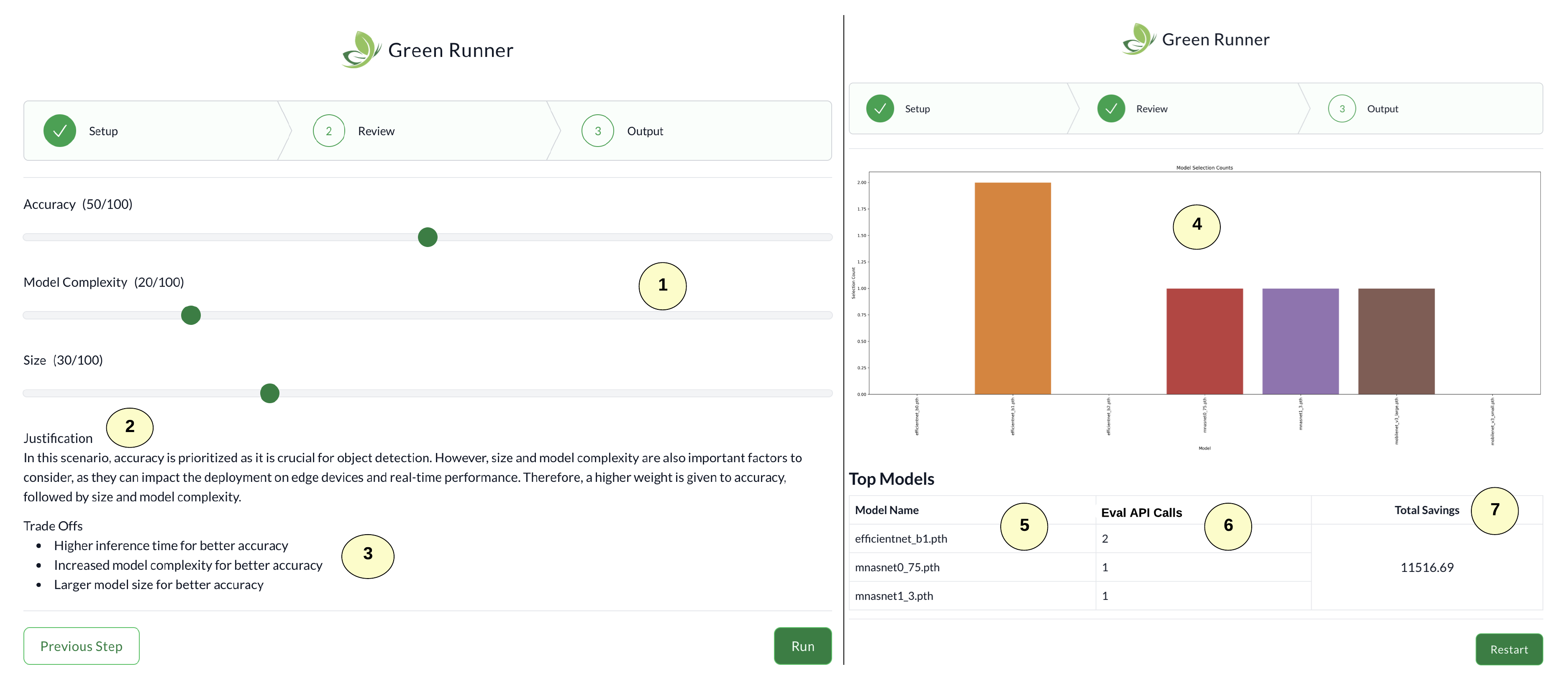}
    \caption{\toolname{}'s user interface displaying the configuration of an experiment and the analysis report.}
    \label{fig:usage}
\end{figure*}

\section{Usage example}
To address Tom's challenge, we present an innovative tool designed to optimize model selection while minimizing resource consumption. Tom inputs a simple query like \textit{"Recommend a model for drone-based object detection"} and provides the target dataset. This triggers our GPT-Based reasoning module, which proposes a set of metrics and their importance as shown in \autoref{fig:usage} \numberedcircle{1}. The module also justifies these choices in \numberedcircle{2}, linking them to key use case considerations such as model size, complexity, and performance, detailed in \autoref{sec:motivation}.

Tom can then adjust these metric weights prior to running the experiment \numberedcircle{3}, which will affect the outcome of the resource-efficient experiment engine. After running the experiment, Tom reviews the results on the analysis screen in \numberedcircle{4}, which displays the top models in \numberedcircle{5}, the number of evaluations made per model in \numberedcircle{6}, and the computational savings in \numberedcircle{7} compared to a brute force approach. From here, Tom can select the best model to fine-tune and deploy on his drone.

\section{Preliminary Evaluation}
\label{sec:eval}

To evaluate our approach we proposed two research questions:\textbf{\textit{1. Does \toolname{} find the most suitable model for inferred trade-offs?}} and \textbf{\textit{2. How does \toolname{} compare to other transferability metrics for model selection?}} This evaluates if our approach balances tradeoffs and how it compares to state-of-the-art. We compared \toolname{} to three baselines: 1) benchmark results (results of the model trained on a benchmark dataset), 2) brute force (compare all models on all data points), and 3) a transferability metric from the literature ~\cite{huang2022frustratingly}.  

\textbf{Dataset:} 
In this research, DL components refer to deep learning models selected from a repository. We initially considered the complete collection of image classification models available on PyTorch Hub\footnote{\url{https://pytorch.org/vision/stable/models.html}}, amounting to 80 models. However, due to API issues that prevented the download of 9 models, our experiment proceeded with the remaining 71 models. PyTorch Hub was chosen for its prevalence as an open-source framework in the machine learning community and for the ease of conducting standardized comparisons, as it hosts models pre-trained on the benchmark ImageNet dataset \cite{deng2009imagenet}.
\begin{table*}[h]
\centering
\caption{Comparative analysis of i) benchmark result, ii) Brute Force against \toolname{}}
\begin{tabular}{|l|l|l|c|r|r|}
\hline
Method & Metric & Top Models & {\begin{tabular}[c]{@{}c@{}}Avg. Performance\\ on Target\end{tabular}} & {\begin{tabular}[c]{@{}r@{}}Avg.\\ Model Size\end{tabular}} & {\begin{tabular}[c]{@{}r@{}}Avg. Model\\ Complexity\end{tabular}} \\ 
\hline
Benchmark result & Accuracy & maxvit\_t & 0.29 & 124.5 MB & 19670 MMAC \\ 
 & Accuracy, Size, Complexity & mobilenet\_v3 & 0.17 & 22 MB & 229 MMAC \\ 
\hline
Brute Force & Accuracy& regnet\_y\_128gf & 0.45 & 2581 MB & 127750 MMAC \\ 
 & Accuracy, Size, Complexity & convex\_net\_small & 0.31 & 114 MB & 4470 MMAC \\ 
\hline
\toolname & Accuracy & regnet\_y\_32gf & 0.32 & 581 MB & 32380 MMAC \\ 
 & Accuracy, Size, Complexity & swin\_v2\_s & 0.30 & 199 MB & 5790 MMAC \\ 
\hline
\end{tabular}
\label{table:Resultanalysis}
\end{table*}

In our experiments, we used the ObjectNet dataset \cite{barbu2019objectnet} as our target for model evaluation. ObjectNet, known for challenging benchmark models, is tailored for testing vision models in realistic settings. For our initial assessment, we randomly selected 200 images from 113 classes corresponding to ImageNet categories to evaluate the models.

\textbf{\toolname{} configuration:} To calculate the metrics and weights for the use case we used a GPT Based Reasoning module (based on GPT-4) using the following prompt \textit{Recommend a model for detecting objects deployed on a drone}. As the response is non-deterministic, we executed the prompt 100 times resulting in average weights for the use case. For the purpose of the experiment we only chose the accuracy, model size and complexity metrics which are: accuracy weight of 0.63, size weight of 0.25, and complexity weight of 0.21 respectively. 

\textbf{Method:} To answer: \textit{1. Does \toolname{} find the most suitable model for inferred trade-offs?}, We performed a comparative analysis using two metric combinations against the benchmark and brute force approaches. The first set of metrics evaluated accuracy alone, while the second also considered model size and complexity, based on suggested weights for the use case. We compared the selected models in terms of on-target accuracy, size (in MB), and complexity (measured in Million Multiply-Accumulate Operations, or MMACs). Given that our approach can recommend varying models per run, we averaged results over 200 iterations, and presented in \autoref{table:Resultanalysis}.

To answer: \textit{2. How does \toolname{} compare to other transferability metrics for model selection?}, we used the TransRate metric proposed in \cite{huang2022frustratingly} and compared the model selected using TransRate to the model selected by \toolname{} 
\begin{table*}[h]
\centering
\caption{Comparative analysis of TransRate~\cite{huang2022frustratingly} against \toolname{}}
\begin{tabular}{|l|l|l|c|r|r|}
\hline
Method & Metric & Top 3 Models & {\begin{tabular}[c]{@{}c@{}}Avg. Accuracy\\ on Target\end{tabular}} & {\begin{tabular}[c]{@{}r@{}}Avg.\\ Model Size\end{tabular}} & {\begin{tabular}[c]{@{}r@{}}Avg. Model\\ Complexity\end{tabular}} \\ 
\hline
\multirow{3}{*}{TransRate \cite{huang2022frustratingly}} & \multirow{3}{*}{N/A} & vgg11 & 0.23 & 531 MB & 7300 MMAC \\ 
 & & vgg\_16bn & 0.17 & 553 MB & 15530 MMAC \\
  & & shufflenet\_v2 & 0.14 & 30 MB & 593 MMAC \\
\hline
\multirow{3}{*}{\toolname} & \multirow{3}{*}{Accuracy, Size, Complexity} & swin\_s & 0.32 & 199 MB & 5790 MMAC \\ 
 & & regnet\_y\_800mf & 0.21 & 26 MB & 851 MMAC \\
 & & efficientnet\_b0 & 0.20 & 22 MB & 401 MMAC \\
\hline
\end{tabular}
\label{table:TransRateVsGreenRunner}
\end{table*}

\textbf{Results:} \textbf{\textit{1. Does \toolname{} find the most suitable model for inferred trade-offs?}}
In our analysis in \autoref{table:Resultanalysis}, we concentrate on the top model ranked by each approach. The benchmark method often chooses models for their high scores (about 0.80) on benchmark datasets, but these models frequently underperform on target datasets, with accuracy dropping to 0.29. When adjusted for size and complexity,  further declines to 0.17, although this results in a smaller model. However, this method risks overlooking models that may offer superior overall performance.

The brute force method, assessing every model against the full dataset, identifies the most accurate model with an accuracy of 0.45 accuracy on the target dataset. However, its resource-intensive nature is a drawback, rendering the best model (2581MB in size and 127750 MMACs in complexity) impractical for resource-constrained applications like drone deployment.

For \toolname, we implemented Thomson Sampling as the selection strategy. Focusing solely on accuracy, our method outperformed the benchmark by selecting a model with 0.32 accuracy on the target dataset, though it did not reach the brute force method's accuracy.

However, when considering a combination of accuracy, size, and complexity, \toolname selected a model with 0.30 accuracy, close to the brute force method's 0.31. The model's reduced size and lower computational needs offset its slight dip in accuracy. This efficiency in model selection aligns \toolname with practical requirements, streamlining the process to identify the most appropriate model for specific use cases.

\textbf{\textit{How does \toolname{} compare to other transferability metric for model selection?}} In this experiment, we compare GreenRunner with TransRate, as outlined in \cite{huang2022frustratingly}, focusing on performance, size, and complexity metrics crucial for drone use cases. TransRate evaluates models based on transferability scores, assessing how a model's learned features align with a specific target task. This method, while adept in gauging feature adaptability, predominantly measures the relative transferability and doesn't consider the tradeoffs for the usecase.

Conversely, GreenRunner employs a broader approach, emphasizing not only transferability but also model size and computational complexity. This assessment is critical in scenarios like drone-based object detection, where resource efficiency is a key consideration. Considering that TransRate's scores are relative, we compare the top three models selected by each method. As demonstrated in \autoref{table:TransRateVsGreenRunner}, GreenRunner effectively identifies models that strike a balance between accuracy, size, and complexity, surpassing TransRate in meeting the operational requirements of drone applications.
\section{Related Work}
To approach the model selection problem recent research has proposed several online and offline approaches \cite{bugliarello2022iglue,you2021logme,ding2022pactran,pandy2022transferability,Huang2021}.
To provide an approach to solve this problem one approach is to dissect models into distinctive building blocks and reassemble them to produce a customised network \cite{yang2022deep}, however reassembling models is complex, and the proposed approach requires significant computational resources and expertise to implement. Research has also proposed platforms which assess the adaptation to a domain task for models in a model zoo \cite{bugliarello2022iglue}. These platforms require specialised knowledge and need empirical evaluation requiring computational resources.

Proposed approaches have also calculated the class similarity between the dataset the models are trained on vs the target dataset for a downstream task \cite{pandy2022transferability}. The assumption being a high class similarity between datasets produces similar performance between the models.
Research has also produced lightweight tools to estimate transferability such as \cite{huang2022frustratingly,you2021logme}. These approaches utilise a small amount of data to be passed through the models to assess the potential performance of a target task. However, these approaches require at least one pass of the data through the models to estimate the potential performance of the models on the dataset where resources are utilised to compute the performance on the dataset. The way these approaches differ to our approach is that \toolname{} is able to select an appropriate amount of evaluation data on a per model basis and considers trade-offs rather than just optimising for accuracy.

\section{Conclusion and future work}
\label{sec:conclusion}

\toolname{} addresses a practical challenge that developers encounter when integrating machine learning functionality into software. It efficiently selects machine learning models for specific use cases, surpassing traditional benchmark and brute force methods in terms of both optimality and efficiency. Benchmark methods often fall short, as they rely on source data accuracy, which might not reflect the target data performance. brute force methods, while accurate, consume significant resources in evaluating all models. \toolname{} balances high accuracy with manageable model sizes and complexity by leveraging metrics and weights tailored to each use case. This expedites the model selection process and significantly reduces computational resources, thus minimizing resource consumption and environmental impact and contributing to sustainable DL software development.

Our current work focuses on only one part of the DL component selection i.e. choosing appropriate pre-trained models from repositories for a downstream task. Looking ahead, we aim to expand our scope to include model fine-tuning and the selection of machine learning API web services. These areas align with our framework and reward function but will necessitate distinct multi-armed bandit (MAB) strategies and user interfaces. Further developments will explore enabling the GPT-based Reasoning Module to recommend a broader range of metrics beyond accuracy, size, and complexity. This expansion is planned to be seamlessly integrated into our existing system but will call for adjustments to the reward function and strategies for addressing newly important metrics as suggested by the LLM.

\bibliographystyle{ACM-Reference-Format}
\bibliography{references.bib}

\end{document}